\documentclass[referee]{aa} 

\usepackage{graphicx}
\usepackage{txfonts}
\bibpunct{(}{)}{;}{a}{}{,} 
\begin{document}

   \title{Departure from the constant-period ephemeris for the transiting exoplanet WASP-12~b\thanks{Partly based on (1) data collected with the Nordic Optical Telescope, operated on the island of La Palma jointly by Denmark, Finland, Iceland, Norway, and Sweden, in the Spanish Observatorio del Roque de los Muchachos of the Instituto de Astrof\'{\i}sica de Canarias, (2) observations made at the Centro Astron\'omico Hispano Alem\'an (CAHA), operated jointly by the Max-Planck Institut f\"ur Astronomie and the Instituto de Astrof\'{\i}sica de Andaluc\'{\i}a (CSIC), and (3) data collected with telescopes at the Rozhen National Astronomical Observatory.}}

  
   \author{G.~Maciejewski\inst{\ref{inst1}},
          D.~Dimitrov\inst{\ref{inst2}},
          M.~Fern\'andez\inst{\ref{inst3}},
          A.~Sota\inst{\ref{inst3}},
          G.~Nowak\inst{\ref{inst4},\ref{inst5}},
          J.~Ohlert\inst{\ref{inst6},\ref{inst7}},
          G.~Nikolov\inst{\ref{inst2}},
          {\L}.~Bukowiecki\inst{\ref{inst1}},
          T.~C.~Hinse\inst{8},
          E.~Pall\'e\inst{\ref{inst4},\ref{inst5}},
          B.~Tingley\inst{\ref{inst4},\ref{inst5},\ref{inst9}},
          D.~Kjurkchieva\inst{\ref{inst10}},
          J.~W.~Lee\inst{8},
          \and
          C.-U.~Lee\inst{8}
    }

   \institute{Centre for Astronomy, Faculty of Physics, Astronomy and Informatics, 
             Nicolaus Copernicus University, Grudziadzka 5, 87-100 Torun, Poland, 
              \email{gmac@umk.pl}\label{inst1}
         \and
             Institute of Astronomy, Bulgarian Academy of Sciences, 72 Tsarigradsko Chausse Blvd., 1784 Sofia, Bulgaria\label{inst2}
         \and
             Instituto de Astrof\'isica de Andaluc\'ia (IAA-CSIC), Glorieta de la Astronom\'ia 3, 18008 Granada, Spain\label{inst3}
         \and
             Instituto de Astrof\'isica de Canarias, C/ v\'ia L\'actea, s/n, E-38205 La Laguna, Tenerife, Spain\label{inst4}
         \and
             Departamento de Astrof\'isica, Universidad de La Laguna, Av. Astrof\'isico Francisco S\'anchez, s/n, E-38206 La Laguna, Tenerife, Spain\label{inst5}
         \and
             Michael Adrian Observatorium, Astronomie Stiftung Trebur, 65428 Trebur, Germany\label{inst6}
         \and 
             University of Applied Sciences, Technische Hochschule Mittelhessen, 61169 Friedberg, Germany\label{inst7}
         \and
             Korea Astronomy \& Space Science Institute (KASI), 305-348 Daejeon, Republic of Korea\label{inst8}
         \and
             Stellar Astrophysics Centre, Institut for Fysik og Astronomi, {\AA}arhus Universitet, Ny Munkegade 120, 8000 {\AA}arhus C, Denmark\label{inst9}
                      \and
             Department of Physics, Shumen University, 9700 Shumen, Bulgaria\label{inst10}
    }


  \abstract
   {Most hot Jupiters are expected to spiral in towards their host stars due to transfering of the angular momentum of the orbital motion to the stellar spin. Their orbits can also precess due to planet-star interactions. Calculations show that both effects could be detected for the very-hot exoplanet WASP-12~b using the method of precise transit timing over a timespan of the order of 10 yr. We acquired new precise light curves for 29 transits of WASP-12~b, spannning 4 observing seasons from November 2012 to February 2016. New mid-transit times, together with literature ones, were used to refine the transit ephemeris and analyse the timing residuals. We find that the transit times of WASP-12~b do not follow a linear ephemeris with a 5 sigma confidence level. They may be approximated with a quadratic ephemeris that gives a rate of change in the orbital period of $(-2.56 \pm 0.40) \times 10^{-2}$ s~yr$^{-1}$. The tidal quality parameter of the host star was found to be equal to $2.5 \times 10^5$ that is comparable to theoretical predictions for Sun-like stars. We also consider a model, in which the observed timing residuals are interpreted as a result of the apsidal precession. We find, however, that this model is statistically less probable than the orbital decay.}


   \maketitle
%

\section{Introduction}

With its orbital period of about 26 hours, the transiting planet WASP-12~b \citep{2009ApJ...693.1920H} belongs to a group of hot Jupiters on the tightest orbits. It has a mass of $M_{\rm{b}}=1.39\pm0.19$ $M_{\rm{Jup}}$ \citep{2014ApJ...785..126K} and a radius $R_{\rm{b}}=1.90\pm0.09$ $R_{\rm{Jup}}$ \citep{2013AA...551A.108M} that results in a mean density of only 20\% that of Jupiter. The planet is inflated filling $\sim$60\% of its Roche lobe \citep{2010Natur.463.1054L,2011AJ....141...59B}. 

The planet shape departs from spherical symmetry, and both bodies in the system -- the planet and the host star -- raise mutual tides. The nonspherical-mass component of the gravitational field results in precession of the orbit \citep[e.g.][]{2009ApJ...698.1778R}. Such apsidal rotation could be observed through precise timing of transits for non-zero orbital eccentricities. The total apsidal precession is a sum of components due to tidal bulges, rotation bulges, and relativistic effects. With a precession period of 18 yr, WASP-12~b was found to be a promising candidate for detecting apsidal precession, mainly produced by tides risen on the planet \citep{2009ApJ...698.1778R,2011A&A...535A.116D}. The precession rate may be used to determine the second-order Love number, which is related to the planet's internal density profile. 

Planets on short-period orbits are expected to be unstable to tidal dissipation and finally spiral in towards the host star due to transfering of the angular momentum of the orbital motion through tidal dissipation inside the star \citep[e.g.][]{2009ApJ...692L...9L,2016ApJ...816...18E}. The rate of this orbital decay can help determine the efficiency of the dissipation of tides. The decaying orbital period is expected to be observed through transit timing. For some planets, the cumulative shift in transit times may be of order of $10^2$ s after 10 years \citep{2014MNRAS.440.1470B}. Tentative detections of the orbital decay were reported for systems OGLE-TR-113 and WASP-43 \citep{2010ApJ...721.1829A,2014ApJ...781..116B}, but have not been confirmed by further observations \citep{2016MNRAS.455.1334H,2016AJ....151...17J}.

In this study, we analyse new light curves for WASP-12~b's transits, and reanalyze some literature ones, in order to detect any longtime variations in the orbital period that may be attributed to any of the two mechanisms mentioned above.

\section{Observations and data processing}

We acquired 31 complete light curves for 29 transit between November 2012 and February 2016 using seven telescopes:
\begin{itemize}
\item the 2.56 Nordic Optical Telescope (NOT) at the Observatorio del Roque de los Muchachos, La Palma (Spain) with the ALFOSC instrument in spectroscopic mode;
\item the 2.2 m reflector (CA) at the Calar Alto Astronomical Observatory (Spain) with CAFOS in imaging mode;
\item the 2.0 m Ritchey-Chr\'etien-Coud\'e Telescope (ROZ) at the National Astronomical Observatory Rozhen (Bulgaria), equipped with a Roper Scientific VersArray 1300B CCD camera;
\item the 1.8 m Bohyunsan Optical Telescope (BOAO) at the Bohyunsan Optical Astronomy Observatory (South Korea), equipped with a 4k CCD imaging instrument;
\item the 1.5 m Ritchey-Chr\'etien Telescope (OSN) at the Sierra Nevada Observatory (Spain) with a Roper Scientific VersArray 2048B CCD camera;
\item the 1.2 m Trebur 1-meter Telescope (TRE) at the Michael Adrian Observatory in Trebur (Germany), equipped with an SBIG STL-6303 CCD camera;
\item the 0.6 m Cassegrain Telescope (PIW) at the Center for Astronomy of the Nicolaus Copernicus University in Piwnice (Poland), equipped with an SBIG STL-1001 CCD camera.
\end{itemize}
One transit was observed simultaneously with two telescopes, and the another one was observed with a single telescope in two bands. Most of the data were acquired through $R$-band filters. To achieve a higher transit timing precission, some data were acquired without any filter. The list of observing runs is presented in Table~\ref{Tab.Obs}.

\begin{figure}
  \centering
  \includegraphics[width=9cm]{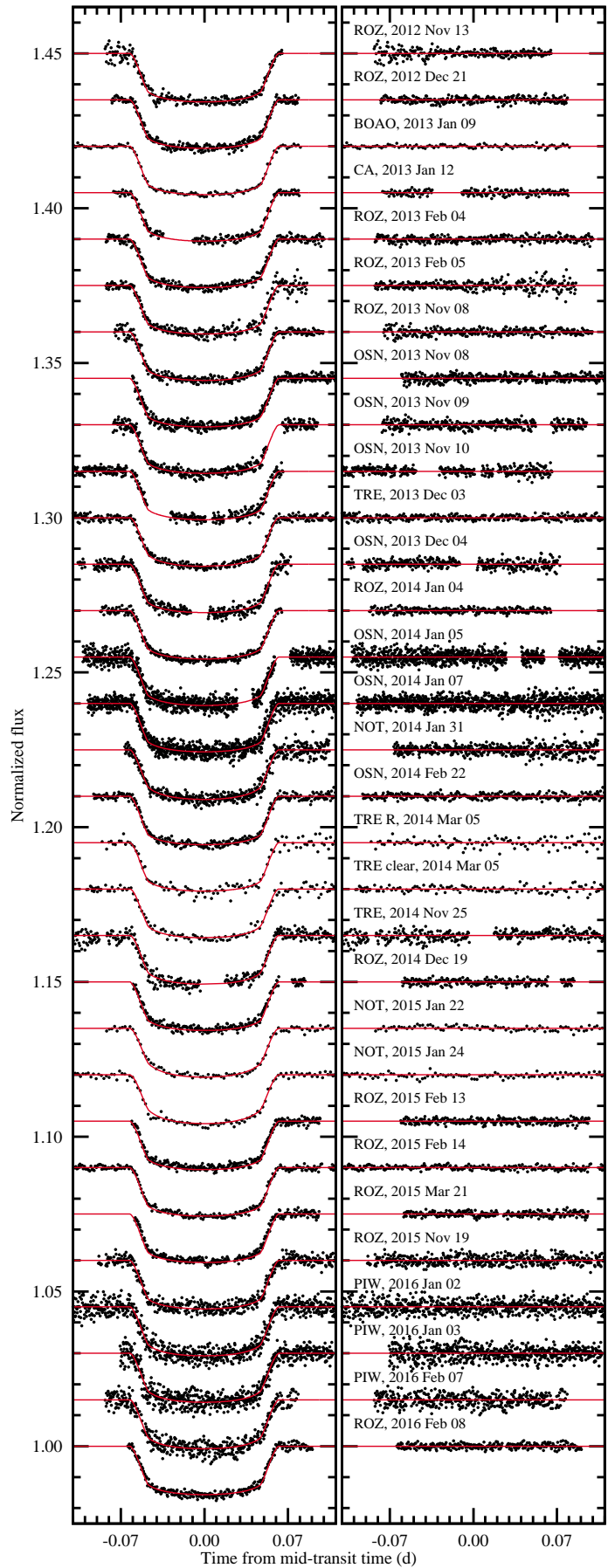}
  \caption{\textit{Left:} new transit light curves for WASP-12~b, sorted by observation date. \textit{Right:} the residuals from the transit model adopted from \citet{2013AA...551A.108M}.}
  \label{Fig:01}
\end{figure}

The data reduction was performed with the AstroImageJ package\footnote{http://www.astro.louisville.edu/software/astroimagej} \citep[\textsc{AIJ},][]{2016arXiv160102622C} following a standard procedure including de-biasing (or dark current removal) and flat-fielding. Photometric time series were obtained with differential aperture photometry. The aperture radius was allowed to vary to compensate for variable seeing. A set of comparison stars was optimized to achieve the lowest photometric scatter and to minimize trends caused by the differential atmospheric extinction. The light curves were examined for linear trends against airmass, position on the matrix, time, and seeing. The fluxes were normalized to unity for out-of-transit brightness. Timestamps were converted to barycentric Julian dates in barycentric dynamical time ($\rm{BJD_{TDB}}$).

The same procedure was applied to re-reduce data acquired with the 2.2 m telescope at Calar Alto, 2.0 m telescope at Rozhen, 1.8 m BOAO, and 1.2 m telescope in Trebur, published in \citet{2011AA...528A..65M} and \citet{2013AA...551A.108M}. The re-reduction improved the photometric quality of the light curves by up to 15\%. 

The light curves from the NOT/ALFOSC were generated from series of low resolution spectra obtained with a custom-built slit of a width of 40\arcsec. Bias and flat-field corrections, extraction of spectra and corresponding calibration arcs, and wavelength calibration based on He and Ne lamps were made using an \textsc{IRAF} script dedicated to NOT/ALFOSC long-slit data. On 2014 Jan 31 a grism \#10, which covers the spectral range $3300-10550$ \AA, was used and light curve was constructed from the whole spectral range. On 2015 Jan 22 and 24 we used a grism \#4, which covers the spectral range $3200-9100$ \AA. The light curves were constructed from the $4950-6050$ \AA  ~spectral range, which corresponds to a photometric $V$-band filter.

\begin{table}
\caption{New mid-transit times for individual epochs.} 
\label{Tab.Res}      
\centering                  
\begin{tabular}{l c c l}      
\hline\hline                
Epoch & $T_{\rm{mid}}$ (BJD$_{\rm{TDB}}$ 2450000+) & $N_{\rm{lc}}$ & Telescope\\
\hline 
1591 & $6245.42729^{+0.00033}_{-0.00033}$ & 1 & ROZ \\
1625 & $6282.53584^{+0.00030}_{-0.00030}$ & 1 & ROZ \\
1643 & $6302.18179^{+0.00046}_{-0.00046}$ & 1 & BOAO \\
1646 & $6305.45536^{+0.00024}_{-0.00026}$ & 1 & CA \\
1667 & $6328.37556^{+0.00027}_{-0.00025}$ & 1 & ROZ \\
1668 & $6329.46733^{+0.00029}_{-0.00029}$ & 1 & ROZ \\
1920 & $6604.50489^{+0.00021}_{-0.00020}$ & 2 & ROZ, OSN \\
1921 & $6605.59624^{+0.00030}_{-0.00030}$ & 1 & OSN  \\
1922 & $6606.68760^{+0.00033}_{-0.00034}$ & 1 & OSN  \\
1943 & $6629.60726^{+0.00019}_{-0.00019}$ & 1 & TRE  \\
1944 & $6630.69917^{+0.00043}_{-0.00043}$ & 1 & OSN  \\
1973 & $6662.35014^{+0.00019}_{-0.00018}$ & 1 & ROZ  \\
1974 & $6663.44136^{+0.00019}_{-0.00019}$ & 1 & OSN  \\
1975 & $6664.53256^{+0.00031}_{-0.00032}$ & 1 & OSN  \\
1997 & $6688.54384^{+0.00040}_{-0.00041}$ & 1 & NOT  \\
2018 & $6711.46415^{+0.00025}_{-0.00026}$ & 1 & OSN  \\
2028 & $6722.37807^{+0.00046}_{-0.00047}$ & 2 & TRE  \\
2270 & $6986.50195^{+0.00043}_{-0.00042}$ & 1 & TRE \\
2292 & $7010.51298^{+0.00039}_{-0.00039}$ & 1 & ROZ \\
2324 & $7045.43831^{+0.00046}_{-0.00049}$  & 1 & NOT \\
2325 & $7046.53019^{+0.00049}_{-0.00047}$  & 1 & NOT \\
2344 & $7067.26715^{+0.00022}_{-0.00023}$  & 1 & ROZ \\
2345 & $7068.35834^{+0.00020}_{-0.00021}$ & 1 & ROZ \\
2377 & $7103.28423^{+0.00031}_{-0.00030}$ & 1 & ROZ \\
2599 & $7345.57867^{+0.00042}_{-0.00040}$ & 1 & ROZ \\
2640 & $7390.32708^{+0.00033}_{-0.00034}$ & 1 & PIW \\
2641 & $7391.41818^{+0.00033}_{-0.00032}$ & 1 & PIW \\
2673 & $7426.34324^{+0.00055}_{-0.00052}$ & 1 & PIW \\
2674 & $7427.43496^{+0.00023}_{-0.00022}$ & 1 & ROZ \\
\hline                                   
\end{tabular}
\tablefoot{Epoch is the transit number from the initial ephemeris given in \citet{2009ApJ...693.1920H}. $T_{\rm{mid}}$ is mid-transit time. $N_{\rm{lc}}$ is the number of individual light curves used simultaneously for a given epoch.}
\end{table}

The set of new light curves was enhanced with photometric time series that are available in the literature. To limit the sample to the most reliable data, we considered only complete transit light curves that were acquired with telescopes with mirrors greater than 1 m. In addition to data from \citet{2011AA...528A..65M} and \citet{2013AA...551A.108M}, we qualified photometric timeseries from \citet{2013MNRAS.434..661C}, \citet{2011AJ....141..179C}, \citet{2012ApJ...747...82C}, and \citet{2014ApJ...791...36S}.

The Transit Analysis Package \citep{2012AdAst2012E..30G} was employed to determine mid-transit times and their uncertainties for individual epochs. The transit parameters such as the orbital inclination, scaled semi-major axis, planetary-to-stellar radii ratio, and coefficients of the quadratic limb darkening law were taken from \citet{2013AA...551A.108M} for $R$-band data and linearly interpolated from tables of \citet{2011A&A...529A..75C} for the remaining bands. During the fitting procedure, the parameters were allowed to vary under Gaussian penalty determined by their uncertainties. The mid-transit time, airmass slope, and flux offsets were the free parameters. The median values of marginalized posteriori probability distributions of the 10 Markov Chain Monte Carlo chains with $10^5$ steps each and the 15.9 and 85.1 percentile values of these distributions were taken as the best-fitting parameters and upper and lower 1\,$\sigma$ errors, respectively.

Mid-transit times for the new light curves, which are shown in Fig.~\ref{Fig:01}, are given in Table~\ref{Tab.Res}. Mid-transit times redetermined from literature data are presented in Table~\ref{Tab.TTimes}. To extend the time covered by the observations, we also used the mid-transit time for epoch 0 from \citet{2009ApJ...693.1920H}. It was obtained from a global fit, so it represents an averaged value for the early epochs.

\section{Results}\label{Sec.Res}

\begin{figure}
  \centering
  \includegraphics[width=12cm]{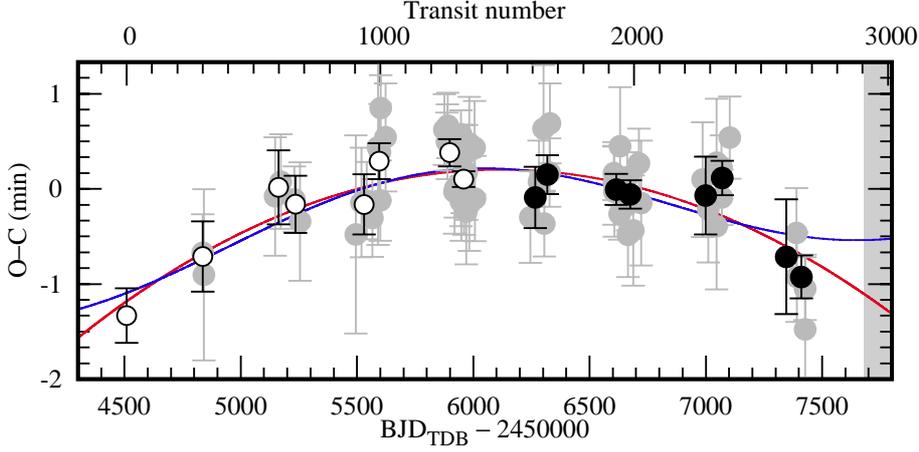}
  \caption{Transit timing residuals from the linear ephemeris. The black open and filled circles denote literature and new data, respectively. Each point represents a half-season everage of individual observations, which are marked with gray dots. The continuous line sketches a quadratic ephemeris discussed in Sect.~\ref{Sec.Res}. The dashed line illustrates a periodic signal discussed in Sect.~\ref{Sec.Dis}. The grayed area shows next observing season, in which new data are expected to distinguish between both models.}
  \label{Fig:02}
\end{figure}

The transit times were used to refine the transit ephemeris and the timing residuals were examined for any long-term variations. Individual timing residuals were binned into halve-season bins. Since complete transits of WASP-12~b are observable from the ground from October to March, the first bin of each observing season included data acquired between October and December and the second one took observations acquired between January and March. Uncertainties of individual mid-transit times were taken as weights, and the weighted standard deviations were calculated as errors. A fit of a linear ephemeris was found to be unsatisfying with the reduced  $\chi$-squared ($\chi^2_{\rm{red}}$) equal to 4.02. This value corresponds to a $p$-value of $5.3 \times 10^{-7}$ that allows us to reject the null hypothesis with a 5 sigma (99.9999\%) conficence level. A Lomb-Scargle periodogram of the timing residuals reveals a long-term ($\sim$3300 d) signal, which is comparable to the timespan covered by the observations. A quadratic ephemeris in the form of
\begin{equation}
  T_{\rm{mid}}= T_0 + P_{\rm{b}} E + \frac{1}{2} {\delta P_{\rm{b}}} E^2,
\end{equation}
where $E$ is the transit number from the cycle zero epoch $T_0$ and $\delta P_{\rm{b}}$ is the change in the orbital period between succeeding transits, yields a much better fit with $\chi^2_{\rm{red}} = 1.03$. We obtained $T_0=2454508.97696\pm0.00016$ BJD$_{\rm{TDB}}$, $P_{\rm{b}}=1.09142162\pm0.00000021$ d, and $\delta P_{\rm{b}}=(-8.9\pm1.4)\times 10^{-10}$ days epoch$^{-2}$. The latter quantity translates into the short-term rate of change in the orbital period ${\dot P}_{\rm{b}} = \delta P_{\rm{b}} / P_{\rm{b}} = (-2.56 \pm 0.40) \times 10^{-2}$ s~yr$^{-1}$. The timing residuals from the linear ephemeris together with the quadratic model are plotted in Fig.~\ref{Fig:02}.

\section{Discussion}\label{Sec.Dis}

The negative value of $\delta P_{\rm{b}}$ may be interpreted as evidence of an orbital decay, which is driven by tidal dissipation in the host star. For a star-planet system, in which the total angular momentum $L_{\rm{tot}}$ is conserved but energy is dissipated due to tides, no equilibrium state exists if $L_{\rm{tot}}$ is smaller than the critical angular momentum $L_{\rm{crit}}$ \citep[see Eq.~(2) in][]{2009ApJ...692L...9L}, which is required for the star–planet system to reach a state of dual synchronization. The spin of the star may be a significant, if not dominant, component of $L_{\rm{tot}}$. It depends on the rotational velocity of the star $\omega_*$, which remains roughly constrained for the  WASP-12 star by spectral observations \citep{2012ApJ...757...18A}. For the WASP-12 system, we obtained $L_{\rm{tot}}/L_{\rm{crit}}=0.4-0.7$, depending on the rotation period of the star. Since this ratio is obviously smaller than 1, the planet will unavoidable spiral inwards\footnote{In calculations we used the normalized moments of inertia (NMoI) equal to 0.04 for the star. The quantity was interpolated from tables of \citet{1989A&AS...81...37C}. For the planet, we used Jupiter's NMoI of 0.26 from \citet{2011Icar..216..440H}.}.  Following Eq.~(5) in \citet{2009ApJ...692L...9L}, we find a relatively short in-spiral time of order of $10^6$ yr that is very short compared to the age of the system of $\sim$$2 \times 10^9$ yr \citep{2009ApJ...693.1920H}.

The observed rate of the orbital decay may be used to determine the stellar tidal quality parameter $Q'_*$. This quantity is the ratio of the stellar tidal quality factor $Q$, which characterizes a body's response to tides, to the second-order stellar tidal Love number $k_2$. Following \citet{2014ApJ...781..116B}, we adopted Eq.~(3) of \citet{2009ApJ...692L...9L} for synchronous planetary rotation and negligible eccentricity and obliquity
\begin{equation}
  Q'_*= 9 P_{\rm{b}} {\dot P}_{\rm{b}}^{-1} \frac{M_{\rm{b}}}{M_*} {\left(\frac{R_*}{a_{\rm{b}}} \right)}^5 {\left( {\omega}_* - \frac{2 \pi}{P_{\rm{b}}} \right)}.
\end{equation}
We obtained $Q'_*$ of order of $2.5 \times 10^5$, which is of the same order as $Q'_* = 4.3 \times 10^5$ calculated from models of \citet{2016ApJ...816...18E} for solar-type host stars.

Alternatively, the observed departure from the linear ephemeris may be a part of the 3300-d periodic signal induced by star-planet tidal interactions. Using Eq.~(86) of \citet{2012CeMDA.113..169M}, we find the time-scale of the rotation of the pericenter due to tides risen in the planet as a response on stellar gravity to be of order of $10^1$ yr. This value could correspond to the observed signal\footnote{Other contributions to the rotation of the pericenter can be neglected. The time scales of the rotation due to rotational deformation of the star and the planet are of order of $10^4$ yr. For the tidal deformation of the star and relativistic effects, the time-scales are of order of $10^3$ yr.}. To explore this possibility, we employed the Systemic software \citep[ver.~2.182,][]{2009PASP..121.1016M}. We used the transit timing dataset enhanced with occultation times from \citet{2011ApJ...727..125C}, \citet{2011AJ....141...30C}, \citet{2012ApJ...760..140C}, and \citet{2015ApJ...802...28C}. We also included a homogenous set of precise radial velocity (RV) measurements from \citet{2014ApJ...785..126K}. This  Doppler time series, free of any year-to-year instrumental trends, was acquired with the Keck/HIRES instrument between December 2009 and December 2013. The orbital period, planetary mass, mean anomaly for a given epoch, eccentricity $e_{\rm{b}}$, longitude of periastron for a given epoch, and periastron precession rate $\dot \omega$ were allowed to be fitted. The Levenbarg-Marquardt algorithm was used to find the best-fitting model with the Keplerian approach. The parameter uncertainties were estimated as median absolute deviations from the bootstrap run of $10^5$ trails. We obtained $e_{\rm{b}}=0.00110\pm0.00036$ and $\dot \omega = 0.095 \pm 0.020$ degrees per day, which corresponds to the period of the periastron precession $\tau_{\omega} = 10.4 \pm 2.2$ yr.

The periastron precession with the very small value of $e_{\rm{b}}$ would have a marginal effect on transit parameters directly determined from light curves. In particular, the range of variations in transit duration is expected to be 2.7 s, much smaller than typical transit duration uncertainties of  1-3 min.

In Fig.~\ref{Fig:02}, we also show the best-fitting sinusoidal signal with the period equal to $\tau_{\omega}$. With $\chi^2_{\rm{red}} = 1.46$, the goodness of the fit is noticeable worse compared the quadratic model. The Bayesian information criterion \citep[BIC,][]{1978AnnSt...6...461E} 
\begin{equation}
  \rm{BIC} = {\chi}^2 + k \ln N,
\end{equation}
where $k$ is the number of fitted parameters and $N$ is the number of data points, also favors the quadratic model with $\rm{BIC}  = 21.8$ over the periodic model with $\rm{BIC} = 28.6$ with a probability ratio of $e^{\Delta \rm{BIC} /2} = 29$. We note that observations in the upcoming season 2016/2017, which is represented by the grayed area in Fig.~\ref{Fig:02}, are expected to definitely distinguish between both models.

\section{Conclusions}

Our new precise observations, spread over 4 years, show that mid-transit times of the WASP-12~b planet do not follow a linear ephemeris. This phenomenon may be interpreted as the result of orbital decay, periastron precession due to planetary tides, or a combination of both effects. The statistics formally favors the orbital decay scenario. The tidal quality parameter for the host star $Q'_*$ was found to be slightly lower than the theoretical predictions. In turn, the periastron precession model is consistent with theoretical predictions and places tight constraints on the orbital eccentricity of the planet. It could be used to precisely determine the Love number of the planet that is related to its internal structure. Further precise timing observations are expected to provide evidence in favor of one of the two scenarios. This will lead to better understanding of the properties of either the host star or the planet. 

\begin{acknowledgements}
We would like to thank NOT and Calar Alto staff for their help during observing runs. We are grateful to Dr.~Chris Copperwheat and Dr.~Nick Cowan for making the light curves available for us. GM acknowledges funding from the European Community's Seventh Framework Programme (FP7/2007-2013) under grant agreement number RG226604 (OPTICON). GM and GN acknowledge the financial support from the Polish Ministry of Science and Higher Education through the Iuventus Plus grant and IP2011 031971. MF was supported by the Spanish grant AYA2014-54348-C3-1-R. TCH, JWL \& CUL acknowledge travel support from KASI grant number 2013-9-400-00 and astronomical observations were carried out during a KRCF Young Scientist Research Fellowship Program. DK acknowledges the financial support from Shumen University, project RD 08-81. A part of this paper is the result of the exchange and joint research project {\em Spectral and photometric studies of variable stars} between the Polish and Bulgarian Academies of Sciences. The data presented here were obtained in part with ALFOSC, which is provided by the Instituto de Astrof\'{\i}sica de Andaluc\'{\i}a (IAA) under a joint agreement with the University of Copenhagen and NOTSA.
\end{acknowledgements}


\bibliographystyle{aa} 
\bibliography{w12rev} 

\appendix

\section{Suplementary materials}

Table~\ref{Tab.Obs} presents details of new light curves acquired for WASP-12~b's transits. Table~\ref{Tab.TTimes} shows mid-transit times based on the literature that were used in our study. 

\begin{table}[h]
\caption{Details on new light curves reported in this paper.} 
\label{Tab.Obs}      
\centering                  
\begin{tabular}{l c c c c c }      
\hline\hline                
Date UT & Epoch & Telescope &  Filter  & $\Gamma$ & $pnr$\\
\hline 
2012 Nov 13 & 1591 & ROZ & $R_{\rm{C}}$ & 1.22 & 1.05 \\
2012 Dec 21 & 1625 & ROZ & $R_{\rm{C}}$ & 1.22 & 0.72 \\
2013 Jan 09 & 1643 & BOAO & $R_{\rm{B}}$ & 0.38 & 0.95 \\
2013 Jan 12 & 1646 & CA & $R_{\rm{C}}$ & 1.22 & 0.72 \\
2013 Feb 04 & 1667 & ROZ & $R_{\rm{C}}$ & 1.22 & 0.74 \\
2013 Feb 05 & 1668 & ROZ & $R_{\rm{C}}$ & 1.22 & 1.17 \\
2013 Nov 08 & 1920 & ROZ & $R_{\rm{C}}$ & 1.22 & 0.82 \\
                     &          & OSN & $R_{\rm{C}}$ & 1.71 & 0.72 \\
2013 Nov 09 & 1921 & OSN & $R_{\rm{C}}$ & 1.69 & 0.80 \\
2013 Nov 10 & 1922 & OSN & $R_{\rm{C}}$ & 1.71 & 0.76 \\
2013 Dec 03 & 1943 & TRE & none & 1.03 & 0.72 \\
2013 Dec 04 & 1944 & OSN & $R_{\rm{C}}$ & 1.71 & 0.86 \\
2014 Jan 04 & 1973 & ROZ & $R_{\rm{C}}$ & 1.22 & 0.61 \\
2014 Jan 05 & 1974 & OSN & $R_{\rm{C}}$ & 3.75 & 0.80 \\
2014 Jan 07 & 1975 & OSN & $R_{\rm{C}}$ & 3.99 & 0.81\\
2014 Jan 31 & 1997 & NOT & none & 1.99 & 0.97 \\
2014 Feb 22 & 2018 & OSN & $R_{\rm{C}}$ & 1.09 & 0.74\\
2014 Mar 05 & 2028 & TRE & $R_{\rm{B}}$ & 0.39 & 2.09\\
                     &         & TRE & none & 0.39 & 1.52 \\
2014 Nov 25 & 2270 & TRE & none & 1.25 & 1.24 \\
2014 Dec 19 & 2292 & ROZ & $R_{\rm{C}}$ & 1.22 & 0.67 \\
2015 Jan 22 & 2324 & NOT & $V$ & 0.32 & 1.17 \\
2015 Jan 24 & 2325 & NOT & $V$ & 0.32 & 1.36 \\
2015 Feb 13 & 2344 & ROZ & $R_{\rm{C}}$ & 1.22 & 0.65 \\
2015 Feb 14 & 2345 & ROZ & $R_{\rm{C}}$ & 0.76 & 0.73\\
2015 Mar 21 & 2377 & ROZ & $R_{\rm{C}}$ & 1.22 & 0.63 \\
2015 Nov 19 & 2599 & ROZ & $R_{\rm{C}}$ & 1.76 & 1.00 \\
2016 Jan 02 & 2640 & PIW & none & 2.62 & 1.20 \\
2016 Jan 03 & 2641 & PIW & none & 2.40 & 1.18 \\
2016 Feb 07 & 2673 & PIW & none & 2.01 & 1.24 \\
2016 Feb 08 & 2674 & ROZ & $R_{\rm{C}}$ & 1.22 & 0.69 \\
\hline                                   
\end{tabular}
\tablefoot{Date UT is given for mid-transit time. Epoch is the transit number from the initial ephemeris given in \citet{2009ApJ...693.1920H}. $R_{\rm{C}}$ and $R_{\rm{B}}$ denote Cousins and Bessel $R$-band filters, respectively. $\Gamma$ is the median number of exposures per minute. $pnr$ is the photometric scatter in $10^{-3}$ normalized flux per minute of observation adopted from \citet{2011AJ....142...84F}.}
\end{table}

\begin{table}
\caption{Mid-transit times based on literature data.} 
\label{Tab.TTimes}      
\centering                  
\begin{tabular}{r c c}      
\hline\hline                
Epoch & $T_{\rm{mid}}$ & Data source\\
  & (BJD$_{\rm{TDB}}$ 2450000+) &  \\
\hline 
0 & $4508.97685^{+0.00020}_{-0.00020}$ & 1 \\
300 & $4836.40340^{+0.00028}_{-0.00028}$ & 2 \\
304 & $4840.76893^{+0.00062}_{-0.00060}$ & 3 \\
585 & $5147.45861^{+0.00043}_{-0.00042}$ & 4 \\
608 & $5172.56138^{+0.00036}_{-0.00035}$ & 3 \\
661 & $5230.40653^{+0.00024}_{-0.00024}$ & 5 \\ 
683 & $5254.41761^{+0.00043}_{-0.00042}$ & 5 \\
903 & $5494.52999^{+0.00072}_{-0.00074}$ & 4 \\
925 & $5518.54147^{+0.00040}_{-0.00040}$ & 6 \\
947 & $5542.55273^{+0.00028}_{-0.00029}$ & 4 \\
969 & $5566.56385^{+0.00028}_{-0.00027}$ & 4 \\
991 & $5590.57561^{+0.00068}_{-0.00071}$ & 4 \\
998 & $5598.21552^{+0.00035}_{-0.00035}$ & 4 \\
1000 & $5600.39800^{+0.00029}_{-0.00030}$ & 4 \\
1001 & $5601.49010^{+0.00024}_{-0.00024}$ & 4 \\
1021 & $5623.31829^{+0.00039}_{-0.00039}$ & 4 \\
1253 & $5876.52786^{+0.00027}_{-0.00026}$ & 4 \\
1263 & $5887.44198^{+0.00021}_{-0.00021}$ & 4 \\
1264 & $5888.53340^{+0.00027}_{-0.00027}$ & 4 \\
1266 & $5890.71635^{+0.00024}_{-0.00024}$ & 4 \\
1293 & $5920.18422^{+0.00031}_{-0.00030}$ & 4 \\
1296 & $5923.45850^{+0.00022}_{-0.00021}$ & 4 \\
1317 & $5946.37823^{+0.00018}_{-0.00018}$ & 4 \\
1318 & $5947.47015^{+0.00017}_{-0.00017}$ & 4 \\
1319 & $5948.56112^{+0.00033}_{-0.00034}$ & 4 \\
1322 & $5951.83536^{+0.00011}_{-0.00011}$ & 7 \\
1323 & $5952.92708^{+0.00013}_{-0.00013}$ & 7 \\
1329 & $5959.47543^{+0.00017}_{-0.00017}$ & 4 \\
1330 & $5960.56686^{+0.00032}_{-0.00040}$ & 4 \\
1339 & $5970.38941^{+0.00039}_{-0.00040}$ & 4 \\
1340 & $5971.48111^{+0.00035}_{-0.00035}$ & 4 \\
1350 & $5982.39509^{+0.00034}_{-0.00033}$ & 4 \\
1351 & $5983.48695^{+0.00035}_{-0.00034}$ & 4 \\
1371 & $6005.31533^{+0.00037}_{-0.00034}$ & 4 \\
1372 & $6006.40637^{+0.00031}_{-0.00033}$ & 4 \\
\hline                                   
\end{tabular}
\tablefoot{Epoch is the transit number from the initial ephemeris given in \citet{2009ApJ...693.1920H}. $T_{\rm{mid}}$ is mid-transit time.}
\tablebib{(1)~\citet{2009ApJ...693.1920H}; (2)~\citet{2013MNRAS.434..661C}; (3)~\citet{2011AJ....141..179C}; (4)~\citet{2013AA...551A.108M}; (5)~\citet{2011AA...528A..65M}; (6)~\citet{2012ApJ...747...82C}; (7)~\citet{2014ApJ...791...36S}.}
\end{table}

\end{document}